\documentclass[conference]{IEEEtran}
\usepackage{ulem}
\usepackage{cite}
\usepackage{textcomp}
\usepackage{xcolor}
\usepackage{stfloats}
\usepackage[numbers]{natbib}
\usepackage{lineno,hyperref}
\usepackage{amsmath,amssymb,amsfonts}
\usepackage{algorithmic}
\usepackage{graphicx}
\usepackage{textcomp}
\usepackage{xcolor}
\usepackage{multirow}
\usepackage{array}
\usepackage{colortbl}
\usepackage{verbatim}
\usepackage{url}
\usepackage{float}
\usepackage{tcolorbox}
\usepackage[linesnumbered,lined,boxed,commentsnumbered]{algorithm2e}
\usepackage{booktabs}
\usepackage{amsmath, bm}
\usepackage{listings}
\usepackage{subfigure}
\usepackage{makecell}
\usepackage{multirow}
\usepackage{ulem}
 \usepackage{fancyhdr}

\newcommand{\tool}{ComFormer}

\begin{document}

\title{{\tool}: Code Comment Generation via Transformer and Fusion Method-based Hybrid Code Representation
}

\author{\IEEEauthorblockN{Guang Yang\IEEEauthorrefmark{2},  Xiang Chen\IEEEauthorrefmark{2}\IEEEauthorrefmark{3}\IEEEauthorrefmark{1}, Jinxin Cao\IEEEauthorrefmark{2}, Shuyuan Xu\IEEEauthorrefmark{2}, Zhanqi Cui\IEEEauthorrefmark{3}, Chi Yu\IEEEauthorrefmark{2}, Ke Liu\IEEEauthorrefmark{2}}
\IEEEauthorblockA{\IEEEauthorrefmark{2}\textit{School of Information Science and Technology},
\textit{Nantong University}, China\\
\IEEEauthorrefmark{2}\textit{Computer School},
\textit{Beijing Information Science and Technology University}, China\\
\IEEEauthorrefmark{3}\textit{Key Laboratory of Safety-Critical Software (Nanjing University of Aeronautics and Astronautics)},\\
\textit{Ministry of Industry and Information Technology}, China\\
Email: 1930320014@stmail.ntu.edu.cn, xchencs@ntu.edu.cn, alfred7c@ntu.edu.cn\\ 
1113585141@qq.com, czq@bistu.edu.cn, yc\_struggle@163.com, 806464561@qq.com}
}

% \author{\IEEEauthorblockN{Guang Yang}
% Nantong, China \\
% 1930320014@stmail.ntu.edu.cn}
% \and
% \IEEEauthorblockN{Xiang Chen}
% Nantong, China \\
% xchencs@ntu.edu.cn}
% \and
% \IEEEauthorblockN{Zhanqi Cui}
% Nanjing, China \\
% czq@bistu.edu.cn}
% \and
% \IEEEauthorblockN{Ke Liu}
% Nantong, China \\
% 806464561@qq.com}
% \and
% \IEEEauthorblockN{Chi Yu}
% Nantong, China \\
% yc\_struggle@163.com}
% \and
% \IEEEauthorblockN{Jinxin Cao}
% Nantong, China \\
% alfred7c@ntu.edu.cn}
% }

\maketitle

\begingroup
\renewcommand{\thefootnote}{}
\footnotetext[1]{\IEEEauthorrefmark{1} Xiang Chen is the corresponding author.}
\endgroup

\begin{abstract}

Developers often write low-quality code comments due to the lack of programming experience, which can reduce the efficiency of developers’ program comprehension. Therefore, developers hope that code comment generation tools can be developed to illustrate the functionality and purpose of the code. Recently, researchers mainly model this problem as the neural machine translation problem and tend to use deep learning-based methods. In this study, we propose a novel method 
{\tool} based on Transformer and fusion method-based hybrid code presentation. Moreover, to alleviate OOV (out-of-vocabulary) problem and speed up model training, we further utilize the Byte-BPE algorithm to split identifiers and Sim\_SBT method to perform AST Traversal. We compare {\tool} with seven state-of-the-art baselines from code comment generation and neural machine translation domains. Comparison results show the competitiveness of {\tool} in terms of three performance measures. Moreover, we perform a human study to verify that {\tool} can generate high-quality comments.

\end{abstract}


\section{Introduction}

%background
With the increasing complexity and evolutionary frequency of software projects, the importance of program comprehension is also increasing. A recent study by Xia et al.~\cite{xia2017measuring} showed developers spend 59\% of their time on program comprehension on average during software development and maintenance. 
Therefore, high-quality code comments are critical to improving the efficiency of developers' program comprehension~\cite{he2019understanding}.
However, developers often write low-quality code comments or do not write code comments due to the limited project development budget, lack of programming experience, or insufficient attention to writing code comments.
Although some tools (such as JavaDoc~\cite{kramer1999api} and Doxygen\footnote{http://www.doxygen.org}) can assist in generating code comment templates, these tools still unable to automatically generate content related to the functionality and purpose of the focused code.
If developers manually write code comments, it will be time-consuming and difficult to guarantee the quality of the written comments. Moreover, existing code comments should be updated automatically with the evolution of the related code~\cite{kramer1999api}. Therefore, it is of great significance to design novel methods that can automatically generate high-quality comments after analyzing the focused code.

%related studies
Code comment generation\footnote{This challenging research problem is also called source code summarization in some previous studies~\cite{wan2018improving}\cite{leclair2019neural}} is an active research topic in the current program comprehension research domain.
Research achievements in this research problem can also improve other software engineering tasks (such as software maintenance, code search, and code categorization).
In the early phase, 
most of the studies~\cite{haiduc2010supporting}\cite{haiduc2010use}\cite{sridhara2010towards}\cite{sridhara2011automatically} on code comment generation were based on template-based methods or information retrieval-based methods. 
Recently, most of the studies~\cite{iyer2016summarizing}\cite{hu2018deep}\cite{hu2018summarizing} started to follow an encoder-decoder framework and achieved promising results.

%motivation and novelty of our method
In this study, we propose a novel method {\tool} via Transformer~\cite{vaswani2017attention} and fusion method-based hybrid code representation.
Our method considers Transformer since this deep learning model can achieve better performance than traditional sequence to sequence models in classical natural language processing (NLP) tasks (such as neural machine translation~\cite{vaswani2018tensor2tensor}\cite{raganato2018analysis} and software engineering~\cite{cao2021automated}).
Moreover, our method also utilizes the hybrid code representation  to effectively learn the semantic of the code since this representation can extract both lexical-level and syntactic-level information from the code, respectively. In the hybrid code representation, we not only consider sequential tokens of source code (i.e., lexical level of code) but also utilize AST (abstract syntax tree) information by our proposed Sim\_SBT method (i.e., syntactic level of code). Moreover, we also consider three different  methods to fuse this information. Finally, to alleviate the OOV (out-of-vocabulary) problem, we utilize the byte-level Byte-Pair-Encoding algorithm (Byte-BPE)~\cite{wang2020neural} to split identifiers.
To evaluate the effectiveness of our proposed method {\tool}, we conduct experimental studies on a large-scale code corpus, which contains 485,812 pairs. Each pair includes a Java method and corresponding code comment. This corpus was gathered by 
Hu et al. ~\cite{hu2020deep}. They performed a set of data cleaning steps to ensure the high quality of this corpus. Until now, this corpus has been widely used as the experimental subject in previous  code comment generation studies~\cite{hu2020deep}\cite{hu2018deep}\cite{kang2019assessing}\cite{wang2020reinforcement}\cite{zhang2020retrieval}\cite{wei2019code}.

We design empirical studies and perform human studies to verify the effectiveness of our proposed method. 
We first compare {\tool} with four state-of-the-art baselines from code comment generation (i.e., DeepCom~\cite{hu2018deep}, Hybrid-DeepCom~\cite{hu2020deep}, Transformer~\cite{ahmad2020transformer}, CodePtr~\cite{2021NIU}) and three baselines from neural machine translation (i.e., seq2seq models~\cite{sutskever2014sequence} with/without attention mechanism~\cite{bahdanau2015neural} and GPT-2~\cite{radford2019language})  in terms of three performance measures (i.e., BLEU, METEOR, and ROUGE-L), which are classical measures in previous code comment generation studies. Empirical results show {\tool} can improve the performance when compared with these state-of-the-art baseline methods.
Second, after comparing three fusion methods (i.e., Jointly Encoder, Shared Encoder, and Single Encoder) to combine code lexical information and AST syntactic information, we find {\tool} with Single Encoder can achieve the best performance. 
Third, We perform a human study to verify the effectiveness of {\tool}. In our human study, we compare the comments generated by {\tool} with the comments generated by Hybrid-DeepCom~\cite{hu2020deep}, which has the best performance among the chosen baselines. The results of our human study also show the competitiveness of {\tool}. 
% Finally, we discuss the effectiveness of {\tool} for code with different lengths when compared to Hybrid-DeepCom.

%main contributions

To our best knowledge, the main contributions of our study can be summarized as follows:

\begin{itemize}
\item  We propose a novel code comment generation method {\tool} based on the Transformer and the fusion method-based hybrid code representation.  Instead of the copy mechanism, we mitigate the OOV problem through the Byte-BPE algorithm and vocabulary sharing.  Then we propose a simplification version of the SBT algorithm (i.e., Sim\_SBT) to traverse the structural information of the AST, which can speed up model training. Finally, we consider three different methods for fusing lexical and
syntactical information of the code.

\item We evaluate the performance of our proposed method {\tool} on a large-scale code corpus, which contains 485,812 Java methods and corresponding code comments. The experimental results show that {\tool} is more effective than seven state-of-the-art baselines from both the code comment generation domain and the neural machine translation domain in terms of three performance measures. Moreover, we further conduct a human study to verify the effectiveness of {\tool}.

\item We share our source code, trained models, and used code corpus in the GitHub repository\footnote{\url{https://github.com/NTDXYG/ComFormer}}, which can facilitate the replication of {\tool} and encourage other researchers to make a comparison with {\tool}.
\end{itemize}

\noindent\textbf{Paper organization.} The rest of the paper is organized as follows. 
Section~\ref{sec:ground} presents the background and related work of our study. 
Section~\ref{sec:method} shows the framework of our proposed method {\tool} and details of key components in {\tool}. 
Section~\ref{sec:setup}  shows the experiment setup.
Section~\ref{sec:results} analyzes our empirical results. 
Section~\ref{sec:discussion} performs a discussion on our proposed method {\tool}.
Section~\ref{sec:threats} discusses potential threats to the validity of our empirical study.
Finally, Section~\ref{sec:conclusion} concludes this paper and shows potential future directions for our study.

\section{Related Work}
\label{sec:ground}

In the early phase, most studies~\cite{sridhara2010towards,sridhara2011generating,sridhara2011automatically,wang2017automatically,mcburney2015automatic,moreno2013automatic,abid2015using,haiduc2010use,haiduc2010supporting,eddy2013evaluating,rodeghero2014improving,rodeghero2015eye,wong2015clocom} used template-based or information retrieval-based methods to generate code comments.
Recently, most of the studies  followed deep learning-based methods (i.e., encoder-decoder framework) and achieved promising results.
Iyer et al.~\cite{iyer2016summarizing} first proposed a method code-NN via an attention-based neural network.
%, which is based on LSTM and attention mechanism in encoder and decoder.
Allamanis et al.~\cite{allamanis2016convolutional} proposed a model in which the encoder uses CNN and attention mechanisms, and the decoder uses GRU. The use of convolution operations helps to detect local time-invariant features and long-range topical attention features.
Zheng et al.~\cite{zheng2017code} proposed a new attention module called Code Attention, which can utilize the domain features (such as symbols and identifiers) of code segments.
Liang and Zhu~\cite{liang2018automatic} used Code-RNN to encode the source code into the vectors, and then they used Code-GRU to decode the vectors to code comments.

% Hu et al.~\cite{hu2018deep} proposed the method DeepCom, which analyzes the structure and semantic information of the Java method through an abstract syntax tree, and then converts the AST into a sequence. In order to better represent the structural information in the AST and ensure that the converted sequence is not ambiguous, they proposed a new AST traversal method SBT (structure-based traversal). SBT encloses the subtree contained by the node in a pair of parentheses. Therefore, the use of brackets can indicate the structure of the AST, and can unambiguously restore the converted sequence to the corresponding AST before the conversion. In addition, in order to further solve the OOV problem, they proposed a new method of representing unknown morphemes. The AST transformed sequence they analyzed contains three types of nodes: terminal nodes, non-terminal nodes and brackets. Wherein morpheme belonging to unknown terminal nodes, so that the method using a general purpose morphemes value used morphemes previously unknown type instead $<UNK>$.

Hu et al.~\cite{hu2018deep} proposed a method DeepCom by analyzing abstract syntax trees (ASTs). To better present the structure of ASTs, they proposed a new structure-based traversal (SBT) method.
Later, Hu et al.~\cite{hu2020deep} further proposed the method Hybrid-DeepCom. This method mainly made some improvements. For example, the identifiers satisfying the camel casing naming convention are split into multiple words.
%The beam search is utilized during code comment generation.
Recently, Kang et al.~\cite{kang2019assessing} analyzed whether using the pre-trained word embedding can improve the model performance. They surprisingly found that using the pre-trained word embedding based on code2vec~\cite{alon2019code2vec} or Glove~\cite{pennington2014glove} does not necessarily improve the performance.

Leclair et al.~\cite{leclair2019neural} proposed a method ast-attendgru, which combines words from code and code structure.
Leclair et al.~\cite{leclair2020improved} then used a Graph neural network (GNN), which can effectively analyze the AST structure to generate code comments.
Wan et al.~\cite{wan2018improving}  proposed the method Hybrid-DRL to alleviate the exposure bias problem.
They input an AST structure and sequential content of code segments into a deep reinforcement learning framework (i.e., actor-critic network).
Then, Wang et al.~\cite{wang2020reinforcement} extended the method Hybrid-DRL.
They used a hierarchical attention network by considering multiple code features, such as type-augmented ASTs and program control flows.

Ahmad et al.~\cite{ahmad2020transformer} used the Transformer model to generate code comments. The Transformer model is a kind of sequence to sequence model based on multi-head self-attention, which can effectively capture long-range dependencies. Specifically, they proposed to combine self-attention and copy attention as the attention mechanism of the model and analyzed the influence of absolute position and pairwise relationship on the performance of the code comment generation.

% Niu et al. propose a new model for code annotation generation, CodePtr, which on the one hand solves the problem of broken code structure by adding a full source code encoder, and on the other hand alleviates the OOV problem by introducing a pointer generation network module.

Chen et al.~\cite{chen2018neural} proposed a neural framework, which allows bidirectional mapping between a code retrieval task and a code comment generation task. Their proposed framework BVAE has two Variational AutoEncoders (VAEs): C-VAE for source code and L-VAE for natural language.
Ye et al.~\cite{ye2020leveraging} exploited the probabilistic correlation between code comment generation task and code generation task via dual learning.
Wei et al.~\cite{wei2019code} also utilized the correlation between code comment generation task and code generation task and proposed a dual training framework.

On the other hand, Hu et al.~\cite{hu2018summarizing} proposed a method TL-CodeSum, which can utilize API knowledge learned in a related task to improve the quality of code comments.
Zhang et al.~\cite{zhang2020retrieval} proposed a retrieval-based neural code comment generation method. This method enhances the model with the most similar code segments retrieved from the training set from syntax and semantics aspects.
Liu et al.~\cite{liu2019neural} utilized the knowledge of the call dependency between the source code and the dependency of codes.
Zhou et al.~\cite{zhou2019augmenting} proposed a method ContextCC, which uses the program analysis to extract context information (i.e., the methods and their dependency).
Haque et al.~\cite{haque2020improved} modeled the file context (i.e., other methods in the same file) of methods, then they used an attention mechanism to find words and concepts to generate comments.

% \noindent\textbf{Commit message generation.}
% Commit messages are natural language descriptions of code changes (i.e., diffs).
% Jiang et al.~\cite{jiang2017automatically} utilized neural machine translation to generate commit messages after analyzing diffs automatically.
% Liu et al.~\cite{liu2018neural} revisited the study of Jiang et al.~\cite{jiang2017automatically}. They found that most of the diffs in the testing set can generate high-quality commit messages by their proposed method~\cite{jiang2017automatically} since these diffs can find similar diffs in the training set. Then they proposed a faster method NNGen based on the nearest neighbor algorithm.
% Xu et al.~\cite{xu2019commit} proposed the method CoDiSum. They first extracted both code structure and code semantics from the diffs. Then they augmented their model via a copying mechanism.
% Liu et al. ~\cite{liu2019automatic} further generate descriptions for pull request. Their method used the sequence to sequence model. Moreover, they considered the pointer generator and used reinforcement learning and a special loss function to further optimize the model performance.

Different from the previous studies, {\tool} is designed based on  Transformer and fusion method-based hybrid code presentation. In this study, we investigate three different methods to fuse lexical-level and syntactic-level code information.
Moreover, we utilize the Byte-BPE algorithm to alleviate the OOV problem and use the Sim\_SBT method to reduce the size of the sequence
generated by the original SBT method~\cite{hu2018deep}, which can speed up model training.

\section{Our Proposed Method {\tool}}
\label{sec:method}

\begin{figure*}[htbp]
	\centering
    \vspace{-1mm}
	\includegraphics[width=0.8\textwidth]{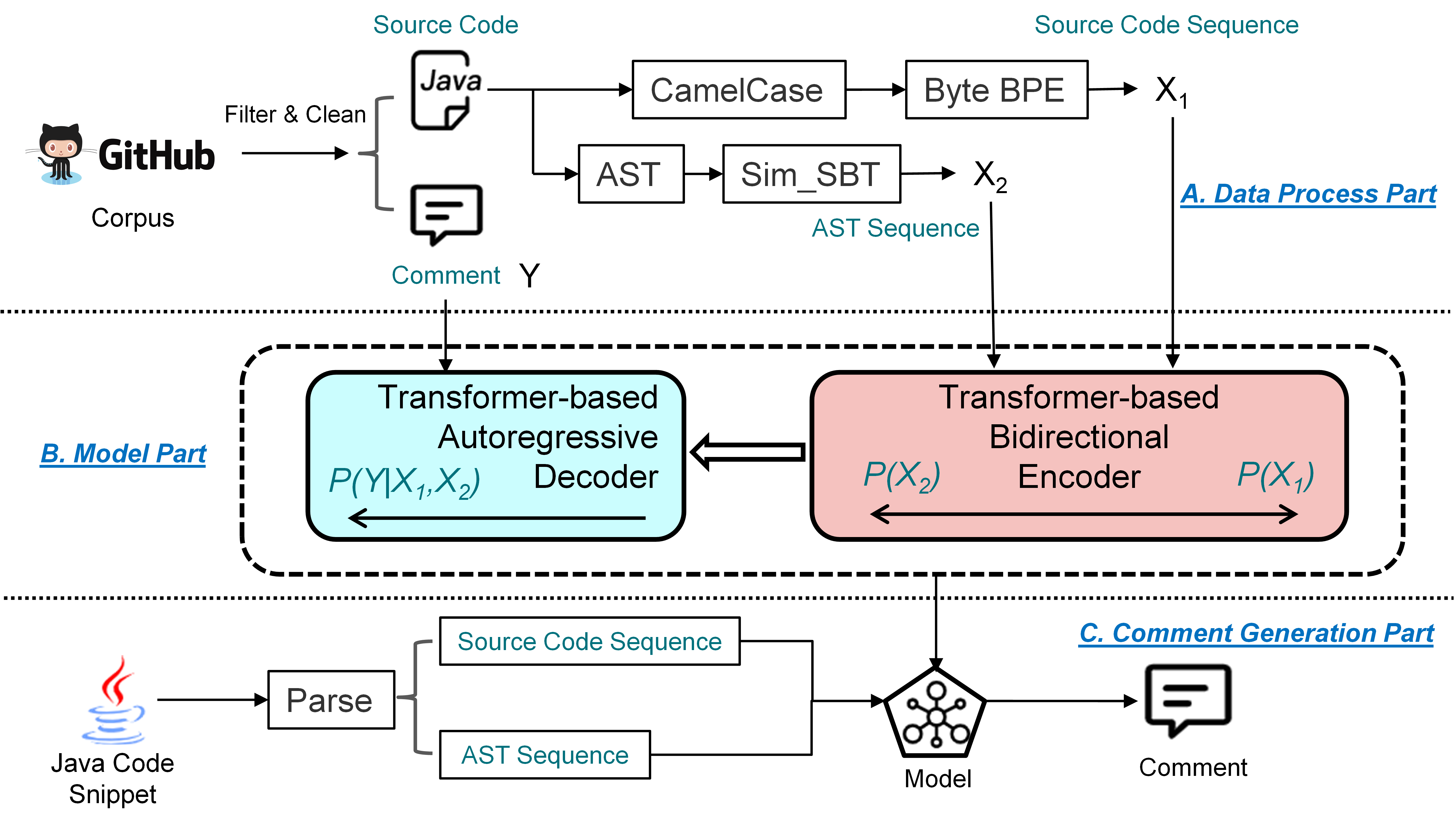}
	\caption{The framework of our proposed method {\tool}}
    \vspace{-1mm}
	\label{fig:approach}
\end{figure*}

Fig.~\ref{fig:approach} shows the framework of {\tool}. In this figure, we can find that {\tool} consists of three parts: data process part, model part, and comment generation part. 
Then we show the details of these three parts.

% \subsection{Framework of {\tool}}

% In particular,
% in the data process part, we first convert the tokens of code into the \textbf{Source Code Sequence}.
% Then we convert the ASTs of the code into the \textbf{AST Sequence} by using simplified version of the SBT~\cite{hu2018deep} method which called Sim\_SBT.
% To alleviate the problem of OOV, 
% we split the identifiers satisfying camel casing or underscore naming convention into multiple words. Then we use Byte-BPE algorithm to split code tokens into sub-tokens. Finally, we share the vocabulary between the encoder and decoder to achieve a similar effect to the pointer network.
% In the model part, we follow the Transformer framework for the construction of the model, and in particular, we make improvements and related experiments for the Encoder. We try to combine lexical-level information (Code Sub-tokens) and syntactic-level information (AST) at the encoder layer, and we propose three different ways of combining them and comparing their effects.
% In the Decoder, we input the information learned by the encoding layer into the decoder for decoding, output the probability of each character generated in the vocabulary, and use the Beam Search algorithm to generate the final code annotations.
% In the comment generation, we parse the source code fragments to obtain the \textbf{Source Code Sequence} and \textbf{AST Sequence}, which are fed into the trained model, which eventually generates the relevant code comments.
% In the following subsections, we will show the details of each steps in our proposed method {\tool}.

\subsection{Data Process Part}

In {\tool}, we consider a hybrid representation of code. For this representation, we not only consider sequential tokens of source code (i.e., lexical level of code) but also utilize AST structure information (i.e., syntactical level of code).

\subsubsection{Constructing Source Code Sequence.} 

We first convert the tokens of code into the sequences.
However, many tokens are identifiers  (such as the class name, the method name, the variable name).
These identifiers are named according to Java's naming convention (i.e., camel casing naming convention). Therefore, most of the identifiers are OOV tokens.
In our study, we first split these identifiers into multiple words, which helps to alleviate the OOV problem and keep more code information. 
For example, the variable name ``SegmentCopy" can be split into two words: "segment" and "copy". The method name ``onDataChanged" can be split into three words: ``on", ``data" and ``changed". The class name ``SecureRandom" can be split into two words: ``secure" and ``random". After splitting the identifiers into multiple words, we then convert all the tokens into lowercase. Finally, we replace the specific numbers and strings with ``$<$num\_$>$" and ``$<$str\_$>$" tags, respectively. 

Second, after performing a more detailed manual analysis on the training set, we find that after splitting the words based only on the camel casing naming convention, there are still a large number of OOV words in the testing set. Most of the current studies~\cite{2021NIU,ahmad2020transformer} alleviated this problem through the copy mechanism by using the pointer network. In our study, we  use the Byte-BPE algorithm~\cite{wang2020neural} to further divide the token of the code into sub-tokens, then combine the vocabulary sharing to solve the OOV problem.
For example, we find the word "forgo" exists in the comments of the test set, which does not appear in the comments of the training set, nor the corresponding source code. In this case, neither the camel casing naming split nor the copy mechanism can solve the problem. However, in the Byte-BPE algorithm, the word ``forgotten" is split into ``for", ``go", ``t", ``ten", so that in The Decoder, it can decode the comments to produce the correct word ``forgo".

\subsubsection{Constructing AST Sequence} 

We first use the javalang tool\footnote{\url{https://pypi.org/project/javalang/}} to convert the Java code into the corresponding AST. Then we use our proposed Sim\_SBT method to generate the traversal sequence of the AST. 
Since the sequences generated by the SBT method~\cite{hu2018deep} may contain redundant information (i.e., many parentheses between type nodes), the sequences generated by SBT traversal are sometimes longer than the source code sequences, which makes it more difficult for the model to learn syntactic information. To alleviate this problem, we propose a new method Sim\_SBT, which can better present the structure of ASTs and keep the sequences unambiguous.

\begin{figure}[htbp]
	\centering
    \vspace{-1mm}
	\includegraphics[width=0.5\textwidth]{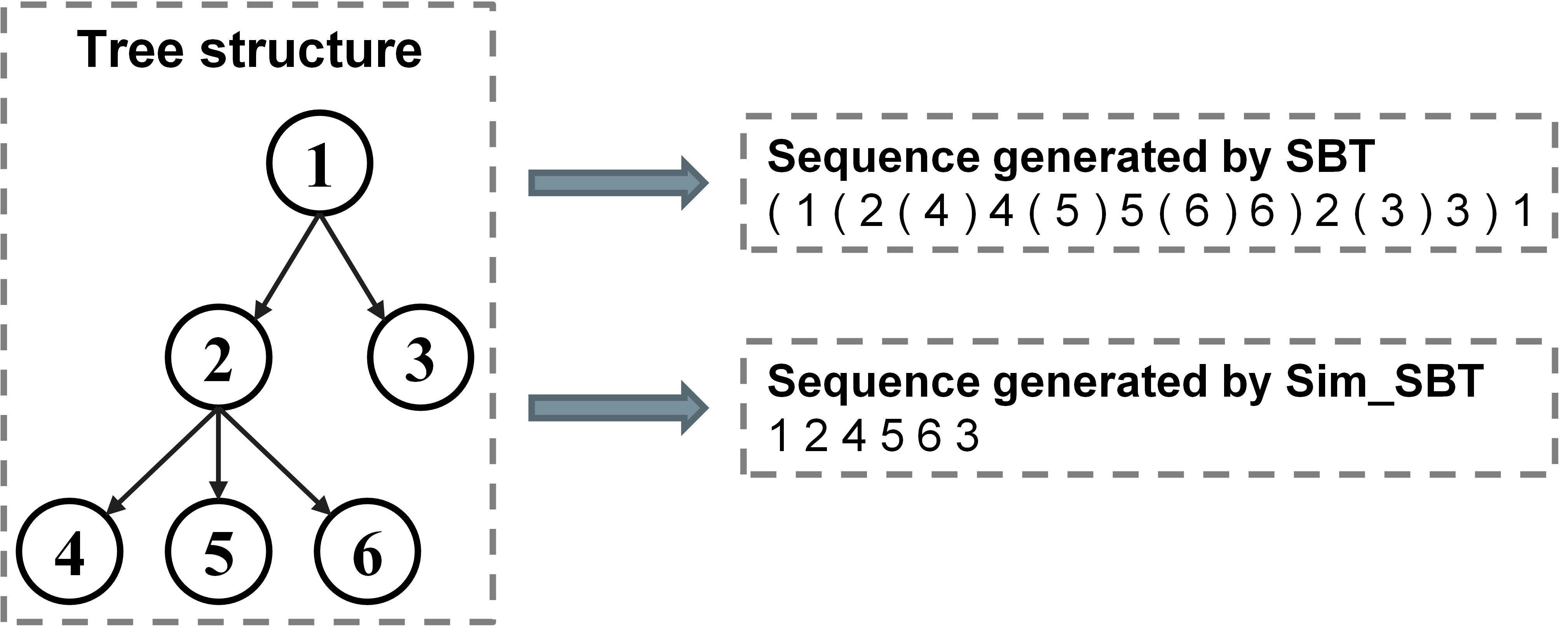}
	\caption{The AST traversal results of the methods SBT and Sim\_SBT}
    \vspace{-1mm}
	\label{fig:SimSBT}
\end{figure}

\begin{figure}[htbp]
	\centering
    \vspace{-1mm}
	\includegraphics[width=0.5\textwidth]{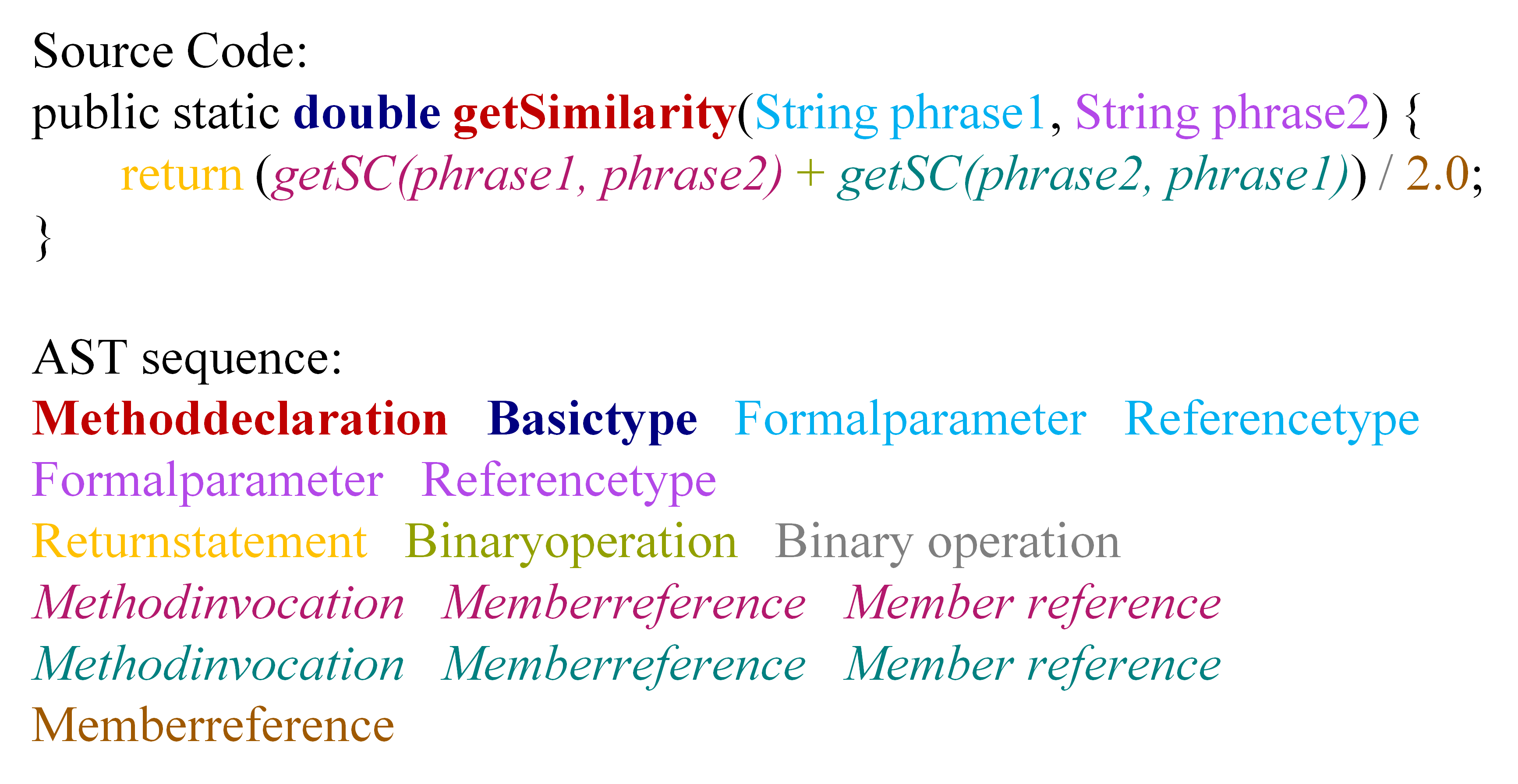}
	\caption{An example of convert an AST to a sequence by using our proposed method  Sim\_SBT.}
    \vspace{-1mm}
	\label{fig:ast}
\end{figure}

The AST traversal results of the methods SBT and Sim\_SBT are shown in Fig.~\ref{fig:SimSBT}. 
In this example, the sequence generated by the original SBT method is too long. Our proposed method Sim\_SBT adopts a prior order traversal in a tree, which has the advantage of reducing the length of the sequence. We use a code example to show the generated sequence by using our proposed method Sim\_SBT in Fig.~\ref{fig:ast}. 
In this figure, the source code token of the same color corresponds to the token of the AST syntax type. We can find  the AST sequence generated by Sim\_SBT is slightly shorter than the source code length, which can effectively reduce the time of model training. 

\subsection{Model Part}

{\tool}  follows the Transformer architecture (i.e., the encoder and the decoder are built  using the self-attentive mechanism). Moreover, {\tool} considers three methods for fusing lexical and
syntactical information of the code at the Encoder.

\subsubsection{Encoder Layer}

In this section, we first introduce Transformer's Encoder and then illustrate three different fusion methods at the Encoder.

The Encoder of Transformer does not attempt to compress the entire source sentence $X = (x_1, \cdots ,x_n)$ into a single context vector $z$. 
Instead it produces a sequence of context vectors $Z = (z_1, \cdots, z_n)$.
First, the tokens of the input are passed through a standard embedding layer. 
Next, as the model has no recurrent, it has no idea about the tokens' order in the sequence. 
This problem is solved by using another embedding layer (i.e., positional embedding layer). 
The positional embedding layer's input is not the token itself but the token position in the sequence.
Notice the input starts with   $<$SOS$>$ (i.e., start of the sequence) token, which is the first token in position 0.
The original implementation of Transformer~\cite{vaswani2017attention}  uses fixed static embeddings and does not learn positional embeddings.
Recently, positional embeddings have been widely used in modern Transformer architectures (such as Bert~\cite{devlin2018bert}). Therefore, our study also uses this positional embedding layer.

The encoded word embeddings are then used as the input to the encoder, which consists of $N$ layers. Each layer contains two sub-layers: (a) a multi-head attention mechanism and (b) a feed-forward network.

A multi-head attention mechanism builds upon scaled dot-product attention, which operates on a
query $Q$, a key $K$, and a value $V$. The original attention calculation uses scaled dot-product for each representation:

\begin{equation}
\text { Attention}(Q, K, V)=\operatorname{softmax}\left(\frac{Q K^{T}}{\sqrt{d_{k}}}\right) V
\end{equation}

Multi-head attention mechanisms obtain $h$ different representations of ($Q$, $K$, $V$). Then concatenate the results, and project the concatenation with a feed-forward layer:

\begin{equation}
\text { head}_{i}=\text { Attention }\left(Q W_{i}^{Q}, K W_{i}^{K}, V W_{i}^{V}\right)
\end{equation}
\begin{equation}
\text { MultiHead}(Q, K, V)=\text {Concat}_{i}\left(\text {head}_{i}\right) W^{O}
\end{equation}
where $W_{i}$ and $W^{O}$ are parameter projection matrices that are learned, and $h$ denotes the number
of heads in the multi-head attention.

The second component of each layer of the Transformer network is a feed-forward network. 
\begin{equation}
\mathrm{FFN}(x)=\max \left(0, x W_{1}+b_{1}\right) W_{2}+b_{2}
\end{equation}

\begin{figure}[htbp]
	\centering
    \vspace{-1mm}
	\includegraphics[width=0.4\textwidth]{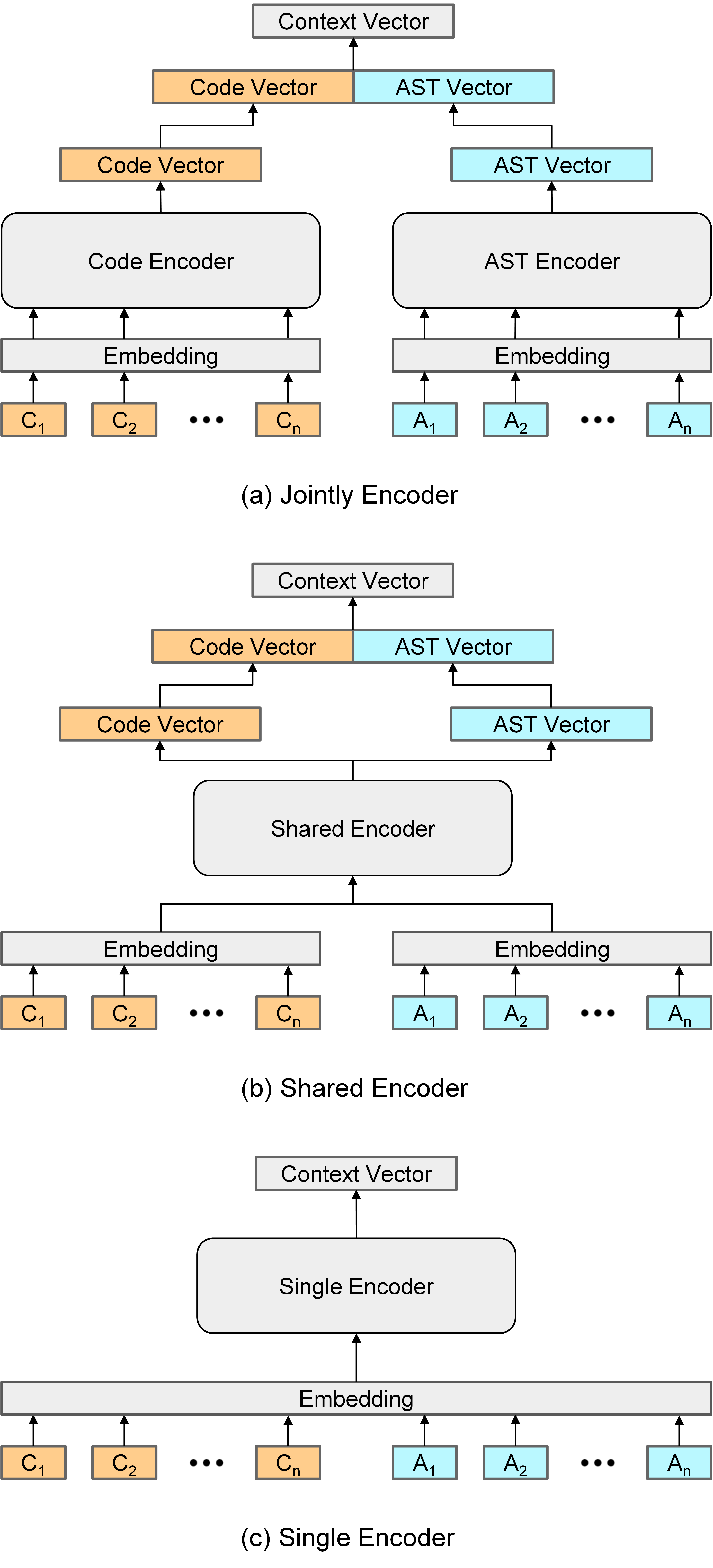}
	\caption{Structure of three different fusion methods in the Encoder}
    \vspace{-1mm}
	\label{fig:encoder}
\end{figure}

Next, we illustrate three different methods (i.e., Jointly Encoder, Shared Encoder, and Single Encoder), which can fuse
lexical and syntactical information of the code at the
Encoder. The structure of these fusing methods can be found in Fig.~\ref{fig:encoder}. Specifically,
\textbf{Jointly Encoder} assumes that AST and source code are two different levels of the input.
Therefore, this method sets up an encoder for the source code sequence (i.e., Code Encoder) and an encoder for the AST Sequence (i.e., AST Encoder), respectively. The Linear layer is activated by the Tanh function to obtain the final matrix of contextual information. 
\textbf{Shared Encoders} considers the effect of having two encoders on the model parameters. This method encodes the source code sequence and the AST Sequence by weight sharing (i.e., using one encoder). Then it switches the two output matrices together, adds a Linear layer, and activates it with the Tanh function to obtain the final matrix of contextual information. 
\textbf{Single Encoder} first splices the source code sequence and the AST sequence. Then, this method proceeds through the word embedding in the encoder afterward, which relies entirely on the positional information encoded in the model for learning lexical and syntactical information. 
%Finally, to confirm the validity of our model improvements, we compared encoders encoded via Source Code only (means  no AST information attached).

\subsubsection{Decoder Layer}

According to the structure of the Transformer, it can be found that the Decoder is  the same as the Encoder. In the beginning, a position vector Positional Encoding was added first, which is the same as the method used in the Encoder. 

Next is the masked multi-head attention, 
the mask represents a mask, which masks certain values so that it has no effect when the parameters are updated. The decoder implements the autoregressive model by means of Mask.
The sequence mask is to make the decoder unable to see future information. That is, for a sequence, at the moment time step is $t$, our decoded output should only depend on the output before time $t$, not the output after $t$. So we need to generate an upper triangular matrix, the values of the upper triangle are all 0. By applying this matrix to each sequence, the information after $t$ can be hidden.

Finally, The combined embeddings are passed through the $N$ decoder layers, along with the encoded source, and the source and target masks. Notice the rest of the layer structure is the same as the Encoder in our method.

\subsection{Comment Generation Part}

Previous studies~\cite{hu2018deep}\cite{hu2018summarizing} showed that generating the text through the maximum probability distribution of the neural networks often yields a low-quality result. 
Recently, most studies~\cite{freitag2017beam}\cite{wiseman2016sequence} resorted to Beam Search~\cite{freitag2017beam}, which can achieve high performance on text generation tasks. 
Therefore, {\tool} uses the Beam Search algorithm to generate code comments.
\section{Experimental Setup}
\label{sec:setup}

In our empirical study, we want to answer the following three research questions (RQs):

\noindent\textbf{RQ1: Can our proposed method {\tool} outperform state-of-the-art baselines for code comment generation in terms of neural machine translation-based measures?}

\noindent\textbf{Motivation.} In this RQ, we want to compare the performance of  {\tool} with the state-of-the-art baselines from both the code comment generation domain and neural machine translation domain  in an automated manner. The main challenge is how to measure the similarity between the comments written by developers and the comments generated by {\tool} and baselines. In this RQ, we consider three performance measures, which have been used in the previous studies on neural machine translation~\cite{lin2004automatic}\cite{guo2018hierarchical} and code comment generation~\cite{hu2018summarizing}\cite{wan2018improving}\cite{hu2020deep}. 

\noindent\textbf{RQ2: Can hybrid code representation improve the performance of our proposed method {\tool}?}

\noindent\textbf{Motivation.} In this RQ, we want to show the effectiveness of fusion Method-based hybrid code representation. Therefore, we want to compare this code representation method with the methods, which only consider code lexical information.
Moreover, we want to compare the performance of different methods for fusing code lexical information and code syntactical information. Then we can select the best fusion method in this study.

\noindent\textbf{RQ3: Can our proposed method {\tool} outperform state-of-the-art  baselines for code comment generation via human study?}

\noindent\textbf{Motivation.} Evaluating the effectiveness of our proposed method in terms of performance measures has the following disadvantages. First, the quality of the comments written by developers can not be guaranteed in some cases. Second, sometimes evaluation based on word similarity is not accurate since two semantic similar code comments may contain different words.  
Therefore, it is necessary to evaluate the effectiveness of our proposed method via human study in a manual way.

% \noindent\textbf{RQ4: How effective is our proposed method {\tool} to source code and comments with different lengths?}

% \noindent\textbf{Motivation.} In RQ1, we mainly evaluate the performance of {\tool} in the coarser granularity (i.e., on the code corpus level). Then in this RQ, we want to evaluate the performance of {\tool} in the finer granularity (i.e., different lengths of Java methods and different lengths of code comments).

\subsection{Code Corpus}

In our empirical study, we choose code corpus\footnote{This corpus can be downloaded from \url{https://github.com/xing-hu/EMSE-DeepCom}} gathered by Hu et al.~\cite{hu2020deep} as our empirical subjects,
since this code corpus have been widely used in previous  studies for code comment generation~\cite{hu2020deep}\cite{hu2018deep}\cite{kang2019assessing}\cite{wang2020reinforcement}\cite{zhang2020retrieval}\cite{wei2019code}.
% The Java methods and their corresponding JavaDoc are collected by Hu et al.~\cite{hu2020deep} from GitHub's Java repositories. To ensure the chosen Java repositories' quality, they only consider the repositories, which have more than ten stars. 

% For the extracted Java methods, they first removed the Java methods without JavaDoc.
% Second, they used the first sentences of the JavaDoc as the target comments since these sentences often describe the functionalities of Java methods according to the guideline of JavaDoc\footnote{\url{http://www.oracle.com/technetwork/articles/java/index-137868.html}}. 
% Then, they removed the Java methods, which their target comments only contain one word.
% Third, they removed the setter/getter methods, constructor, and test methods (i.e., with @Test annotation), since they think generating comments for these kinds of Java methods is trivial.
% Forth, they removed methods with @SmallTest, @LargeTest, and @MediumTest annotations.
% Then they removed the overridden methods since the overridden methods implement the same functionality.
% Finally, after performing the data length distribution analysis, they removed the methods if the methods contain more than 200 tokens or the target comments contain more than 20 tokens.
Table~\ref{tab:statistics} shows the statistical information for code length, SBT length, and comment length. For the above code corpus, 20,000 pairs are  selected to construct the testing set and the validation set. Then the remaining 445,812 pairs are used to construct the training set. This setting is consistent with the experimental setting in the previous studies (such as Hybrid-DeepCom~\cite{hu2020deep}), which can guarantee a fair comparison with the baselines.

\begin{table}[htbp]
 \caption{Statistics of code corpus used in our empirical study} 
 \label{tab:statistics}
 \begin{center}
 \begin{tabular}{cccccc} 
  \toprule 
  \multicolumn{6}{c}{
  \textbf{Statistics for Code Length}} \\ 
    \midrule 
     Avg & Mode & Median & $<100$ & $<150$ & $<200$ \\ 
     55.79 & 11 & 36 & $82.75\%$ & $92.11\%$ & $97.10\%$ \\ 
     \midrule 
  \bottomrule 
  \multicolumn{6}{c}{
  \textbf{Statistics for Comment Length}} \\ 
    \midrule 
     Avg & Mode & Median & $<20$ & $<30$ & $<50$ \\ 
     10.25 & 8 & 9 & $95.69\%$ & $99.99\%$ & $100\%$ \\ 
  \bottomrule 
 \end{tabular}
 \end{center}
\end{table}

% Moreover, in Table~\ref{tab:split}, we also show the unique tokens in code, SBT, and comment sequences.

% \begin{table}[t] 
%  \caption{Split results of code corpus}
%  \label{tab:split}
%  \begin{center}
%  \begin{tabular}{lcl} 
%   \toprule 
%   \textbf{DataSet} & \textbf{Pairs} \\ 
%   \midrule 
%  The training set & 445,812 \\ 
%  The validation set & 20,000 \\ 
%  The testing set& 20,000 \\ 
%  \midrule 
%  Unique tokens in AST & 58 \\
%  Unique tokens in code(previous) & 47,366 \\
%  Unique tokens in comment(previous) & 57,359 \\
%  Unique tokens in code(BPE) & 30,652 \\
%  Unique tokens in comment(BPE) & 32,287 \\
%   \bottomrule 
%  \end{tabular} 
%  \end{center}
% \end{table}

\subsection{Performance Measures}

In our study, we use the performance measures from neural machine translation research to automatically evaluate the quality between the candidate comments (generated by code comment generation methods) and the reference comments (generated by developers). 
The chosen performance measures include BLEU, METEOR, and ROUGE-L. These performance measures have been widely used in previous studies for code comment generation~\cite{hu2018summarizing}\cite{wan2018improving}\cite{hu2020deep}. The details of these performance measures can be found as follows.

\noindent\textbf{BLEU.}
BLEU (Bilingual Evaluation Understudy)~\cite{papineni2002bleu} is the earliest measure used to evaluate the performance of the neural machine translation models. It is used to compare the degree of coincidence of $n$-grams in the candidate text and the reference text. 
In practice, $N$=1$\sim$4 is usually taken, and then the weighted average is performed.
Unigram ($N$ =2) is used to measure word translation accuracy, and high-order $n$-gram is used to measure the fluency of sentence translation.

\noindent\textbf{METEOR.}
METEOR (Metric for Evaluation of Translation with Explicit Ordering)~\cite{banerjee2005meteor} is based on $\mathit{BLEU}$ with some improvements. METEOR is based on the single-precision weighted harmonic mean and the single word recall rate, and its purpose is to solve some inherent defects in the BLEU standard.

\noindent\textbf{ROUGE-L.}
ROUGE-L (Recall-Oriented Understudy for Gisting Evaluation)~\cite{lin2004rouge} calculates the length of the longest common subsequence between the candidate text and the reference text. The longer the length, the higher the score. 

We utilize the implementations provided by nlg-eval library\footnote{\url{https://github.com/Maluuba/nlg-eval}}, which can ensure the implementation correctness of these performance measures.

\subsection{Experimental Settings}

Our proposed method {\tool}  is implemented with PyTorch 1.6.0.
In our study, we choose AdamW as the optimizer and use cross Entropy as the loss function. We set the learning rate  to 0.0005 and set the value of epoch  to 30.

All the experiments run on a computer with an Inter(R) Xeon(R) Silver 4210 CPU and a GeForce RTX3090 GPU with 24 GB memory. The running OS platform is Windows OS. 
\section{Result Analysis}
\label{sec:results}

\subsection{Result Analysis for RQ1}

\noindent\textbf{RQ1: Can our proposed method {\tool} outperform state-of-the-art baselines for code comment generation in terms of neural machine translation-based measures?}

\noindent\textbf{Method.}
In this RQ, we first want to compare our proposed method {\tool} with  Hybrid-DeepCom~\cite{hu2020deep}.
Hybrid-DeepCom used the AST traversal method to represent the code structure information. Then they used the seq2seq model with the attention mechanism to construct the model.
Then, we also choose other four state-of-the-art code comment generation methods (i.e., DeepCom~\cite{hu2018deep},
CodePtr~\cite{2021NIU}, and Transformer~\cite{ahmad2020transformer}) as our baselines.
Later, we choose three baselines from deep learning-based machine translation models.
The first two baselines are traditional seq2seq models~\cite{sutskever2014sequence} with/without attention mechanism~\cite{bahdanau2015neural}. The last baseline is GPT-2~\cite{radford2019language}). GPT2 only uses  the decoder in the Transformer  by large-scale anticipatory learning on tasks (such as machine translation and text summarization).
Finally, to show the competitiveness of our fusion method, we also consider a baseline (i.e., {\tool} without AST), in which the Encoder only considers the code lexical information.
Notice, in this RQ, {\tool} (i.e., {\tool} with AST)  considers the Single Encoder as the fusion method.

For these chosen baselines, we re-use the experimental results of three methods (i.e., DeepCom, Hybrid-DeepCom, and CodePtr) due to the same dataset split setting and re-implement the remaining baselines.

\noindent\textbf{Results.}
The comparison results between {\tool} and the baselines can be found in Table~\ref{tab:RQ1}.
Based on Table~\ref{tab:RQ1}, we can find that
our proposed method {\tool} can outperform all of the baselines. 
In terms of $\mathit{BLEU}$\_1/2/3/4, {\tool} can at least improve its performance by 6.18\%, 9.86\%, 12.76\%, 14.85\% respectively. 
In terms of $\mathit{METEOR}$, {\tool} can improve its performance by 8.20\% at least.
In terms of $\mathit{ROUGE}$-$\mathit{L}$, {\tool} can improve its performance by 4.87\% at least.
Therefore, {\tool} can achieve better performance than the baselines in terms of these performance measures.

\begin{table*}[htbp]
 \caption{The comparison results between our proposed method {\tool} and baseline methods in terms of BLEU, METEOR and Rouge\_L}
 \begin{center}
 \setlength{\tabcolsep}{1mm}{
 \resizebox{0.87\textwidth}{!}{
 \begin{tabular}{ccccccc}
  \toprule
\textbf{METHOD} & \textbf{BLEU\_1(\%)} & \textbf{BLEU\_2(\%)} & \textbf{BLEU\_3(\%)} & \textbf{BLEU\_4(\%)} & \textbf{METEOR(\%)} & \textbf{ROUGE\_L(\%)}  \\
  \midrule
  DeepCom & 49.023  & 44.140 & 38.265 & 35.216  & 25.183 & 52.175\\
  Hybrid-DeepCom  & 54.056  & 45.046 & 40.336 & 37.397  & 27.383 & 54.331\\
  Transformer & 55.624 & 46.295 & 41.574 & 38.692 & 29.056 & 55.263 \\
  CodePtr & 59.506 & 51.107 & 46.386 & 43.371 & 31.382 & 62.761\\
    Seq2Seq & 45.016  & 40.625 & 36.162 & 34.024  & 23.695 & 50.462\\
  Seq2Seq with atten & 46.526  & 41.526 & 37.812 & 35.041  & 24.534 & 51.842\\
  GPT-2 & 47.915  & 41.253 & 37.593 & 35.301 & 26.887 & 53.398\\
\midrule
{\tool} without AST & 59.090 & 51.027 & 46.613 & 43.801 & 31.711 & 60.539 \\
{\tool} with AST & \textbf{62.790} & \textbf{55.283} & \textbf{51.127} & \textbf{48.437} & \textbf{34.182} & \textbf{63.249} \\
  \bottomrule
 \end{tabular} } }
 \end{center}
 \label{tab:RQ1}
\end{table*}

In addition, four code examples with different lengths are selected from the testing set to compare the results generated by {\tool} and baselines. 
% In particular, we choose examples where the comments generated by the {\tool} were identical to the manually written comments and where the comments generated by the baseline method were inconsistent with the manually written comments. We have chosen Hybrid-DeepCom, Codeptr, and {\tool} with/without AST as benchmarks for comparison. The aim is to compare the effectiveness of the different methods in dealing with OOV problems and the effect of adding AST information versus not adding AST information.
The comparison results can be found in Table~\ref{tab: samples of summarizations}. In Case 1, the use of a network of pointers in CodePtr and the use of BPE splitting with vocabulary sharing in {\tool} both generate "cache" words in the comment, which can demonstrate the effectiveness of our method in alleviating the OOV problem. We further verify the competitive nature of our method in Case 2, where the word ``insectwordcategory" does not appear in the source code and {\tool} still generates the comment correctly, while Hybrid-DeepCom and CodePtr only generate $\langle$UNK$\rangle$. As we can find from Case 3 and Case 4, although the comments generated by the baselines are consistent, the comments generated by {\tool} are better after manual analysis. For example, the comment generated in Case 3 explains the reason for doing this separate step, and the comment generated in Case 4 emphasizes the meaning of the if statement.

\begin{table*}[htbp]
 \centering
 \caption{Examples of generated comments by {\tool} and other baselines. These examples cover both long and short code snippets.}
 \label{tab: samples of summarizations}
 \begin{tabular}{c|lp{10cm}}
  \toprule
  Case ID &Example  \\
  \midrule
 
 \multirow{4}{*}{1} &
 \begin{lstlisting}
private void addCachedLegionMemberEx(LegionMemberEx legionMemberEx) {
    this.allCachedLegionMembers.addMemberEx(legionMemberEx);
}

\end{lstlisting}\\
&\textbf{Hybrid-DeepCom:} this method will add a new legion to this container \\
&\textbf{Codeptr:} convenience method to add new member to cache\\
&\textbf{ComFormer without AST:} add legion member to cache\\
&\textbf{ComFormer with AST:} this method will add a new legion member to the cache \\
&\textbf{Human Written:} this method will add a new legion member to the cache \\\midrule

\multirow{4}{*}{2} &
\begin{lstlisting}
public SimpleTextCategory(String sName) {
    super();
    Dictionary = new Dictionary(sName, NUM_);
    Dictionary.AddSubStrings = BOOL_;
    Dictionary.RemoveSubStrings = BOOL_;
    DocumentCount = NUM_;
    Name = sName;
}
\end{lstlisting} \\
&\textbf{Hybrid-DeepCom:} creates a new instance of $\langle$UNK$\rangle$ \\
&\textbf{Codeptr:} creates a new instance of $\langle$UNK$\rangle$\\
&\textbf{ComFormer without AST:} creates a new instance of insectcategory given a category name .\\
&\textbf{ComFormer with AST:} creates a new instance of insectwordcategory \\
&\textbf{Human Written:} creates a new instance of insectwordcategory \\\midrule

\multirow{4}{*}{3} & 
\begin{lstlisting}
public void insertNewAddresses(Dao<AddressBook, Integer> addressBookDao, 
    Dao<AddressItem, Integer> addressItemDao) throws SQLException {
    if (this.nonDaoAddresses == null) {
        return;
    }
    addressBookDao.assignEmptyForeignCollection(this, STR_);
    for (AddressItem addressItem : nonDaoAddresses) {
        addressItem.setAddressBook(this);
        addressItemDao.create(addressItem);
        addresses.add(addressItem);
    }
}

\end{lstlisting}\\
&\textbf{Hybrid-DeepCom:} we need to insert a new contacts into the book.\\
&\textbf{Codeptr:} we have separate ormlite doesn t automatically insert ormlite doesn doesn t automatically insert ormlite doesn doesn t automatically .\\
&\textbf{ComFormer without AST:} insert new contact to database \\
&\textbf{ComFormer with AST:} we have to do this separate step because ormlite doesn t automatically insert children\\ 
&\textbf{Human Written:} we have to do this separate step because ormlite doesn t automatically insert children \\\midrule

\multirow{4}{*}{4} &
\begin{lstlisting}
public static Class<?> findCommonElementType(Collection collection) {
    if (isEmpty(collection)) {
        return null;
    }
    Class<?> candidate = null;
    for (Object val : collection) {
        if (val != null) {
            if (candidate == null) {
                candidate = val.getClass();
            } else if (candidate != val.getClass()) {
                return null;
            }
        }
    }
    return candidate;
}
\end{lstlisting} \\
&\textbf{Hybrid-DeepCom:} finds the common element type for a given collection. \\
&\textbf{Codeptr:} find the common element of the given collection.\\
&\textbf{ComFormer without AST:} find the common element type of the given collection.\\
&\textbf{ComFormer with AST:} find the common element type of the given collection if any. \\ 
&\textbf{Human Written:} find the common element type of the given collection if any. \\

 \bottomrule
\end{tabular}

\end{table*}

% \noindent\textbf{Summary for RQ1:}
% Our proposed method {\tool} can outperform state-of-the-art baselines both from code comment generation domain and neural machine translation domain in terms of three performance measures. Besides, the comments generated by {\tool} can have better quality after analyzing some cases.

\vspace{0.6cm}
\begin{tcolorbox}[width=1.0\linewidth, title={}]
\textbf{Summary for RQ1:}
Our proposed method {\tool} can outperform state-of-the-art baselines both from the code comment generation domain and neural machine translation domain in terms of three performance measures. Besides, the comments generated by {\tool} can have better quality after analyzing some cases.
\end{tcolorbox}

\subsection{Result Analysis for RQ2}

\noindent\textbf{RQ2: Can hybrid code representation improve the performance of our proposed method {\tool}?}

\noindent\textbf{Method.} 
As shown in Fig~\ref{fig:encoder}, we consider three different fusion methods (i.e., Jointly Encoder, Shared Encoder, and Single Encoder) to combine code lexical information and AST syntactical information.

\noindent\textbf{Reults.} 
The comparison results are shown in Table~\ref{tab:RQ2}. 
First, we can find that using these three fusion methods can achieve better performance than {\tool} without AST. This means considering syntactical information from AST can further improve the performance of {\tool}.
Second, among these three fusion methods, Single Encoder can achieve the best performance. This means Single Encoder is best suited for this task.

\vspace{0.6cm}
\begin{tcolorbox}[width=1.0\linewidth, title={}]
\textbf{Summary for RQ2:}
Hybrid code representation can improve the performance of our proposed method {\tool}, while Single Encoder can achieve the best performance.
\end{tcolorbox}
% firstly the three different methods of AST information fusion proposed in this paper all improve the performance of the model compared to the model without the inclusion of AST information, which validates the need to include AST syntax information. Secondly, among the three different methods, we can see that the third method Single Encoder can improve the performance of the model better. 

\begin{table}[htbp]
 \caption{The comparison results between three different fusion methods }
  \renewcommand\arraystretch{1}
 \begin{center}
 \setlength{\tabcolsep}{1mm}{
 \resizebox{0.5\textwidth}{!}{
 \begin{tabular}{cccc}
  \toprule
\textbf{METHOD} &  \textbf{BLEU\_4(\%)} & \textbf{METEOR(\%)} & \textbf{ROUGE\_L(\%)}  \\
  \midrule
  {\tool} without AST & 43.801&  31.711&60.539 \\
Jointly Encoder & 46.301  & 32.925 & 63.012 \\
Shared Encoder & 44.512  & 32.052 & 62.105 \\
Single Encoder & \textbf{48.437} & \textbf{34.182} & \textbf{63.249} \\
  \bottomrule
 \end{tabular} } }
 \end{center}
 \label{tab:RQ2}
\end{table}

\subsection{Result Analysis for RQ3}

\noindent\textbf{RQ3: Can our proposed method {\tool} outperform
state-of-the-art baselines for code comment generation via human study?}

\noindent\textbf{Method.} 
In RQ1, the performance comparison is automatically performed in terms of neural machine translation-based performance measures.
To verify the effectiveness of our proposed methods, we further conduct a  human study.
We recruit two master students majoring in computer science, to perform manual analysis.
Since both of these two master students have rich project development experience, the quality of our human studies can be guaranteed.

Due to the high cost of manually analyzing all the Java methods in the testing set, we use a common sampling method~\cite{singh2013elements} to randomly select at least $MIN$ Java methods and the generated comments from the testing sets. The value of $MIN$ can be determined by the following formula:

\begin{equation}
MIN=\frac{n_{0}}{1+\frac{n_{0}-1}{size}}
\end{equation}
where $n_0$ depends on the selected confidence level and
the desired error margin $n_{0}\left(=\frac{Z^{2} \times 0.25}{e^{2}}\right)$. $Z$ is a confidence
level $z$ score and $e$ is the error margin. $size$ is the number of samples in the testing set. 
For the final manual
analysis, we select $MIN$ examples for the relevant data for the
error margin $e$ = 0.05 at 95\% confidence level (i.e., $MIN =
377$). 

For the 377 selected samples, we show the corresponding source code, the comments generated by {\tool}, and the comments generated by the method Hybrid-DeepCom to master students. 
Notice, these two
master students do not know which method the comment
is generated by, which can guarantee a fair comparison.

Three scores are defined as follows.

\begin{itemize}
\item 1 means that there is no connection between the comment and the code, i.e., the comment does not describe the function and meaning of the corresponding code. We use Low to denote this result.

\item 2 means that the comment is partially related to the code, i.e., it describes part of the function and meaning of the corresponding code. We use Medium to denote this result.

\item 3 means that there is a strong connection between the comment and the code, i.e., the comment correctly describes the function and meaning of the corresponding code. We use High to denote this result.
\end{itemize}

\noindent\textbf{Results.} 
After our human study, we analyze the scoring results of these two master students. The final results are shown in Table~\ref{tab:RQ3}. 
First, we can find {\tool} can generate  a significantly higher proportion of high-quality comments than Hybrid-DeepCom. Then, {\tool} can generate a much lower proportion of low-quality comments than Hybrid-DeepCom. Finally,  {\tool} can achieve a higher score than Hybrid-DeepCom. These results indicate that {\tool} can significantly outperform the baseline Hybrid-DeepCom.

\vspace{0.6cm}
\begin{tcolorbox}[width=1.0\linewidth, title={}]
\textbf{Summary for RQ3:}
Our proposed {\tool} also works better than the baseline method on human study.
\end{tcolorbox}
\begin{table*}[htbp]
 \caption{Manual analysis results on comments generated by {\tool} and Hybrid-DeepCom via human study}
 \label{tab:RQ3}
 \renewcommand\arraystretch{1.5}
 \resizebox{1\textwidth}{!}{
\begin{tabular}{ccccc}
\hline
Student &
  \multicolumn{1}{c}{\begin{tabular}[c]{@{}c@{}}Low\\ {\tool} Hybrid-DeepCom\end{tabular}} &
  \multicolumn{1}{c}{\begin{tabular}[c]{@{}c@{}}Medium\\ {\tool}  Hybrid-DeepCom\end{tabular}} &
  \multicolumn{1}{c}{\begin{tabular}[c]{@{}c@{}}High\\ {\tool}  Hybrid-DeepCom\end{tabular}} &
  \multicolumn{1}{c}{\begin{tabular}[c]{@{}c@{}}Mean\\ {\tool}  Hybrid-DeepCom\end{tabular}} \\ \hline
1 &
  2.39\%   \quad\quad\quad 12.20\% &
  28.12\%  \quad\quad\quad  35.28\% &
  69.49\%  \quad\quad\quad  52.52\% &
  2.67    \quad\quad\quad 2.40 \\
2 &
  4.51\%   \quad\quad\quad 10.34\% &
  29.90\%  \quad\quad\quad  35.01\% &
  65.59\%  \quad\quad\quad  54.65\% &
  2.61   \quad\quad\quad  2.46 \\ \hline
\end{tabular}
}
\end{table*}
\section{Discussions}
\label{sec:discussion}

\begin{figure}
\centering
\subfigure[METEOR scores for different Code lengths]{%
\includegraphics[width=0.5\textwidth]{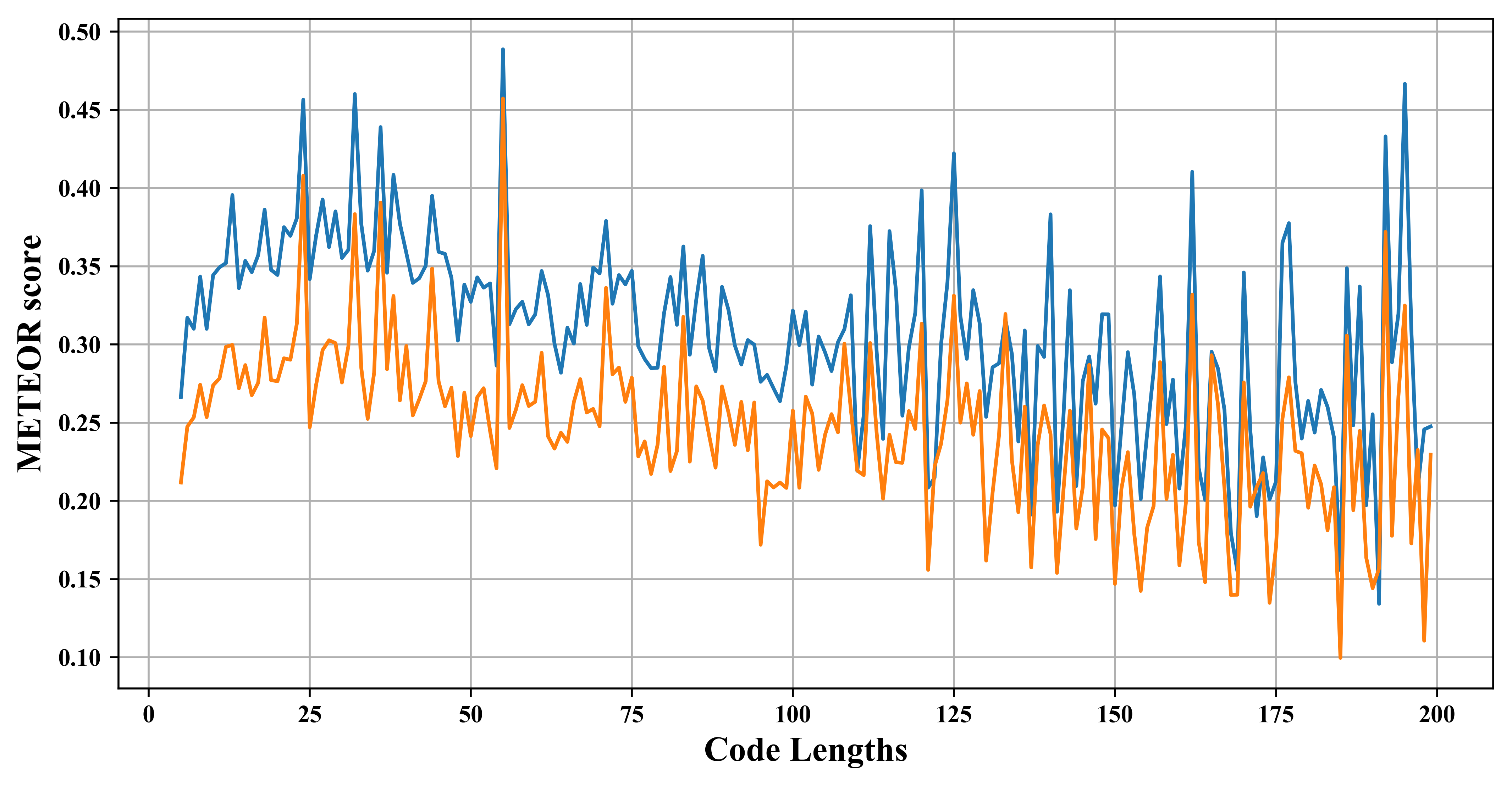}%
}%
\hfill
\subfigure[ROUGE\_L scores for different Code lengths]{%
\includegraphics[width=0.5\textwidth]{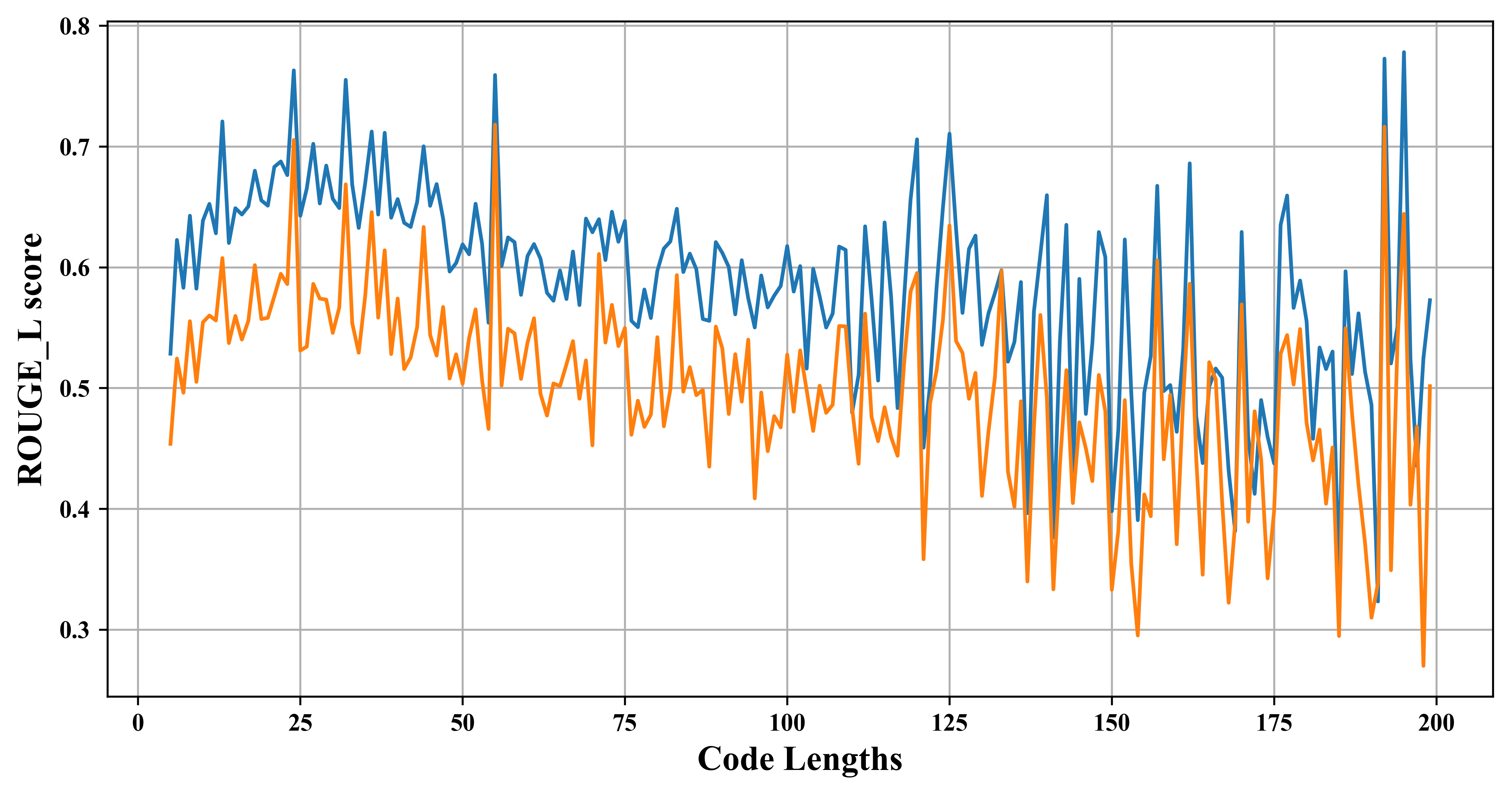}%
}%
\vspace{-1mm}
\caption{Performance comparison between {\tool} and Hybrid-DeepCom by considering different code length in terms of two performance measures, here blue line denotes {\tool} and yellow line denotes Hybrid-DeepCom.}
\label{fig:LEHGTH}
\vspace{-1mm}
\end{figure}

In this section, we aim to analyze the impact of codes' length on the performance of {\tool} and Hybrid-DeepCom. The final results in terms of two performance measures can be found in Fig.~\ref{fig:LEHGTH}. 
As shown in Fig.~\ref{fig:LEHGTH}, the longer the source code length, the lower the average score of METEOR and Rouge\_L. These two methods obtain higher performance when the coding length is between 15$\sim$50.
When the source code length is short, These two methods can learn the full semantics of the source code more easily.
We found that the performance fluctuates significantly when the number of tokens in the source code exceeded 125. Because there are fewer source codes in the corpus, whose length is over 125 and this limits {\tool}'s ability to learn this kind of code.
Overall, 
{\tool} outperformed Hybrid-DeepCom regardless of code length.

\section{Threats to Validity}
\label{sec:threats}

In this section, we mainly discuss potential threats to the validity of our empirical study.

\noindent\textbf{Internal threats.}
The mainly first internal threat is the potential defects in the implementation of our proposed method. To alleviate this threat, we first check code carefully and use mature libraries, such as PyTorch and Transformers\footnote{\url{https://github.com/huggingface/transformers}}.
The second internal threat is the implementation correctness of our chosen baseline methods. To alleviate this threat, we try our best to re-implement their approach according to the original description of these baselines, and our implementation can achieve similar performance reported in their empirical study.

\noindent\textbf{External threats.}
The first external threat is the choice of the corpus. To alleviate this threat, we select a  corpus, which was  provided by Hu et al.~\cite{hu2020deep}. The reasons can be summarized as follows.
First, Java is the most popular programming language, and most of the projects are developed by using Java.
Second, the quality of this code corpus has been improved by Hu et al. by performing data preprocessing. Therefore, this code corpus has also been used in previous studies on code comment generation ~\cite{hu2020deep}\cite{hu2018deep}\cite{kang2019assessing}\cite{wang2020reinforcement}\cite{zhang2020retrieval}\cite{wei2019code}.
In the future, we want to verify the effectiveness of our proposed method for the corpus of other programming languages (such as Python, C\#)~\cite{iyer2016summarizing}.

\noindent\textbf{Construct threats.}
The construct threat in this study is the performance measures used to evaluate our proposed method's performance.
To alleviate these threats, we choose three popular performance measures from the neural machine translation domain. These measures have also been widely used in previous code comment generation studies~\cite{hu2018summarizing}\cite{wan2018improving}\cite{hu2020deep}.
Moreover, we also perform a human study to show the competitiveness of our proposed method.

\noindent\textbf{Conclusion threats.}
The conclusion threat in our study is we do not perform cross-validation in our research.
In our study, the data split on the corpus is consistent with the experimental setting in the previous study for DeepCom~\cite{hu2020deep}.
This can guarantee a fair comparison with the baselines DeepCom, Hybrid-DeepCom, and CodePtr (i.e., the model construction and application on the same training set, validation set, and testing set).
Using cross-validation can comprehensively evaluate our proposed method, since different splits may result in a diverse training set, validation set, and testing set. However, 
this model evaluation method has not been commonly used for neural machine translation experiments due to the high training computational cost.

%\cx{If we use statistical analysis method, we can write this part.}
\section{Conclusion and Future Work}
\label{sec:conclusion}

High-quality code comments are the key to improve the program comprehension efficiency of developers.
Inspired by the latest research advancements in the field of deep learning and program semantic learning,
we propose a novel method {\tool} via Transformer and fusion Method-based hybrid code representation for code comment generation.
In particular, we consider the Transformer to automatically translate the target code to code comment. Moreover, we also use a hybrid code representation (i.e., capture both lexical information and syntactic information) to learn the code semantic effectively. Both empirical studies and human studies verify the effectiveness of our proposed method {\tool}.

In the future,
we first want to evaluate the effectiveness of our proposed method {\tool} by considering other corpus gathered from other programming languages, such as Python, C\#, and SQL query.
Second, we want to use state-of-the-art deep learning methods to improve the performance of our proposed method.
Finally, we also want to design more reasonable performance metrics to  better evaluate the quality of code comments generated by {\tool}.

%\section*{Declaration of Competing Interests}
%	The authors declare that they have no known competing financial interests or personal relationships that could have appeared to influence the work reported in this paper.
%	
%\section*{CRediT Authorship contribution statement}
%\textbf{Guang Yang:} Methodology, Data curation, Software, Writing - original draft.
%\textbf{Xiang Chen:} Methodology, Software, Writing - original draft, Supervision.
%\textbf{Ke Liu:} Data curation, Software, Validation.
%\textbf{Yanxin Jia:} Conceptualization, Writing - review \& editing.

%\section*{Acknowledgment}
%%The authors would like to thank the editors and the anonymous reviewers for their insightful comments and suggestions, which can substantially improve the quality of this work.
%Guang Yang and Xiang Chen have contributed equally to this work and they are co-first authors.
%This work is supported in part by the National Natural Science Foundation of China (Grant nos. 61702041, 61872263 and  61202006), The Open Project of State Key Laboratory of Information Security (Institute of Information Engineering, Chinese Academy of Sciences) under Grant No. 2020-MS-07, and The Open Project of Key Laboratory of Safety-Critical Software for Nanjing University of Aeronautics and Astronautics, Ministry of Industry and Information Technology under Grant No. NJ2020022.

\section*{Acknowledgment}

This work is supported in part by National Natural Science Foundation of
China (Grant nos. 61702041 and  61202006 ), The Open Project of Key Laboratory of Safety-Critical Software for Nanjing University of Aeronautics and Astronautics, Ministry of Industry and Information Technology (Grant No. NJ2020022).

% \section*{Declaration of Competing Interests}
% The authors declare that they have no known competing financial interests or personal relationships that could have appeared to influence the work reported in this paper.

% \section*{CRediT Authorship contribution statement}
% \textbf{Guang Yang:} Data curation, Software.
% \textbf{Xiang Chen:} Methodology, Software, Writing - original draft, Supervision.
% \textbf{Ke Liu:} Data curation, Software, Validation.
% \textbf{Shangqing Liu}: Software, Validation.
% \textbf{Yanxin Jia:} Writing - review \& editing.

%\clearpage
\normalem
\bibliographystyle{IEEEtran}
\bibliography{mylib}
\end{document}